\documentclass[journal]{vgtc}                

\ifpdf
  \pdfoutput=1\relax                   
  \usepackage{graphicx}                
  \DeclareGraphicsExtensions{.pdf,.png,.jpg,.jpeg} 
\else
  \usepackage{graphicx}                
  \DeclareGraphicsExtensions{.eps}     
\fi%

\graphicspath{{figures/}{pictures/}{images/}{./}} 

\usepackage{microtype}                 
\PassOptionsToPackage{warn}{textcomp}  
\usepackage{textcomp}                  
\usepackage{mathptmx}                  
\usepackage{times}                     
\usepackage{cite}                      
\usepackage{tabu}                      
\usepackage{booktabs}                  
\usepackage{balance}
\usepackage{soul}
\usepackage{todonotes}

\usepackage{xcolor,colortbl} 

\onlineid{1082}

\vgtccategory{Research}

\title{PoVRPoint: Authoring Presentations in Mobile Virtual Reality}

\author{Verena Biener, Travis Gesslein, Daniel Schneider, Felix Kawala, Alexander Otte, Per Ola Kristensson, \\Michel Pahud, Eyal Ofek, Cuauhtli Campos, Matjaž Kljun, Klen Čopič Pucihar and Jens Grubert}
\abstract{Virtual Reality (VR) has the potential to support mobile knowledge workers by complementing traditional input devices with a large three-dimensional output space and spatial input. Previous research on supporting VR knowledge work explored domains such as text entry using physical keyboards and spreadsheet interaction using combined pen and touch input. Inspired by such work, this paper probes the VR design space for authoring presentations in mobile settings.
We propose PoVRPoint---a set of tools coupling pen- and touch-based editing of presentations on mobile devices, such as tablets, with the interaction capabilities afforded by VR. We study the utility of extended display space to, for example, assist users in identifying target slides, supporting spatial manipulation of objects on a slide, creating animations, and facilitating arrangements of multiple, possibly occluded shapes or objects. Among other things, our results indicate that 1) the wide field of view afforded by VR results in significantly faster target slide identification times compared to a tablet-only interface for visually salient targets; and 2) the three-dimensional view in VR enables significantly faster object reordering in the presence of occlusion compared to two baseline interfaces. A user study further confirmed that the interaction techniques were found to be usable and enjoyable.
} 


\CCScatlist{ 
 \CCScat{K.6.1}{Management of Computing and Information Systems}%
{Project and People Management}{Life Cycle};
 \CCScat{K.7.m}{The Computing Profession}{Miscellaneous}{Ethics}
}

\teaser{
  \centering
  \includegraphics[width=\linewidth]{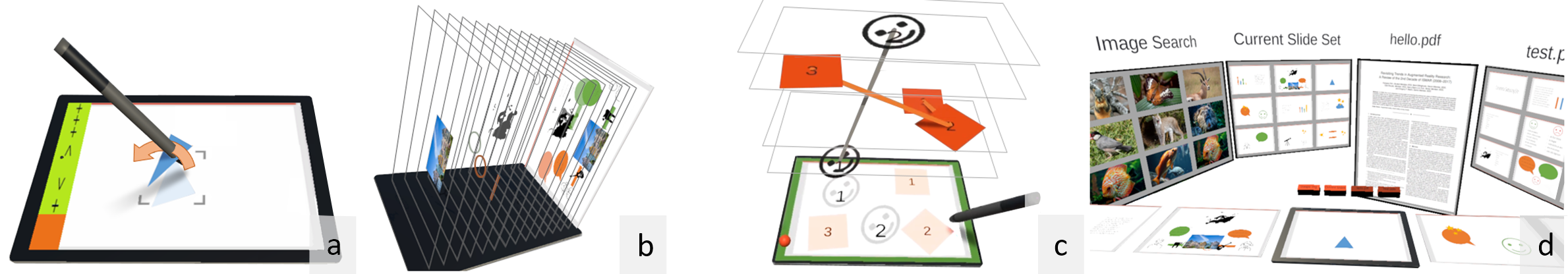}
  \caption{Interaction techniques for authoring presentations in mobile VR. a) 3D object manipulation technique, b) occlusion handling, c) animations with time represented through height, d) working across slides and displaying different content resources}
	\label{fig:teaser}
}

\vgtcinsertpkg

\begin{document}


\firstsection{Introduction}

\maketitle

Using slide-based presentation tools, such as Apple's Keynote and Microsoft's PowerPoint, has widespread use across many sectors, such as education, business and academia. 
In recent years, more people have had the need to work in less than perfect environments, at home, in makeshift offices and on the go. To enable workers to be productive everywhere, even the small space of a middle seat on a coach flight, we need to focus on using small, portable devices such as tablet computers and the small input space where the user may move their hands without interfering with their physical environment.  

Virtual Reality (VR) as an instance of Extended Reality (XR), can benefit users immensely, as it provides a much larger display space than traditional mobile devices, independent of the physical environment and allows rendering of 3D elements above the screen, potentially mitigating effects of a small physical screen.\cite{Biener2020Breaking, Gesslein2020Pen, grubert2018office}. HMDs can also increase privacy and reduce environmental clutter \cite{grubert2018office, schneider2019reconviguration}.
Using VR to support knowledge workers on the go, and therefore in limited spaces, has already been discussed in previous work~\cite{grubert2018office, mcgill2019challenges,  zielasko2019non} and interaction techniques for such scenarios have been developed for multi-screen environments ~\cite{Biener2020Breaking} or spreadsheet applications~ \cite{Gesslein2020Pen}.
This approach is also supported by the advancement of new hardware, as it is already possible to use inside-out tracking HMDs on the go and multiple manufacturers work on light HMDs as a replacement for physical screens. (While our prototype system relies on a stationary tracking system, we expect tracking systems to be available in mobile scenarios in the foreseeable future). Furthermore, new HMDs are designed as monitor replacements, increasing mobility of information workers in the near future \cite{pavanatto2021we, MtRogers2021}, which we hope will increase the popularity of immersive displays for productivity. 

In this work, we focus on exploring how to support the editing process of slide-based presentations using VR in a mobile scenario and present our prototype called \textit{PoVRPoint}. However, common interaction techniques for VR relying on in-air interaction with controllers or hands might not be well-suited for the limited interaction space in mobile settings, especially regarding practicality and social acceptability \cite{mcgill2019challenges}. In addition to the restricted space, fatigue can limit the applicability of in-air interaction \cite{hincapie2014consumed}.
As hand and eye gaze tracking technologies are already being incorporated into commercial HMDs this opens up new interaction possibilities.
Along with mobile devices such as tablets, touch-based interaction can be augmented with spatial interaction above and around the screen~\cite{Biener2020Breaking, Gesslein2020Pen} to facilitate knowledge workers' tasks and potentially improve the overall interaction experience in limited spaces. 
Therefore, we combine a VR HMD with eye-tracking and bimanual pen and touch techniques, where pen-based input can be sensed on and above mobile devices, requiring less interaction space \cite{Gesslein2020Pen}.



We are concentrating on facilitating the authoring process of presentations by expanding typical 2D presentation editing interfaces, such as Microsoft PowerPoint, with an increased 3D output and input space, instead of inherently changing the nature of how presentations are authored today. This makes it possible to leverage familiarity of existing tools and increases compatibility between VR and non-VR working modes.




In this paper we do not present a complete presentation authoring application, but we investigate common tools used in presentation editing that are concerned with both, editing \textit{individual} slides and interacting with presentations across \textit{multiple} elements (multiple slide-sets, PDFs, browsers) and how they can be enhanced using VR.
Specifically, we explored if an extended display space can help users identify target slides (or images) faster, as this is a common task when editing a larger slide-set. In addition, we also investigated how to improve reordering of multiple shapes on a slide in presence of occlusions which can get problematic as the number of objects on a slide increases. We evaluated the performance of these techniques to understand how they can facilitate authoring presentations in VR. For additional concepts and techniques presented in this work, we report a usability study ($n = 18$), confirming the designed interaction techniques are usable and enjoyable. 

In summary, this paper makes four central contributions. First, we present the design and implementation of four VR-based techniques for authoring presentations including manipulating objects, occlusion handling, animations, and working across slides.
Second, we quantify the benefits of using the increased display space of a VR HMD compared to a tablet in a visual search task, e.g., when searching for slides or images. Our results indicate the superiority of VR when the matching task is easy (pre-attentive visual search). However, when the matching difficulty is high (attentive search), VR performs similarly to the tablet. 
Third, we show that the VR-based reordering technique supports significantly faster arranging of occluded objects (such as multiple overlapping shapes on a single slide) compared to the two existing baseline techniques used in PowerPoint.
Fourth, we report that the VR-based techniques have been found usable and enjoyable in a usability study ($n = 18$).

\section{Related Work}
Our work draws on the areas of supporting knowledge workers in XR, pen-based interaction, gaze-based interaction and in-air interaction as well as authoring and presenting in XR.

\subsection{Knowledge Workers in VR}
The use of XR for supporting knowledge work has attracted recent research interest, e.g.,~\cite{grubert2018office, ruvimova2020transport, guo2019mixed, mcgill2020expanding}. 
Early research focused on  projection systems to extend stationary physical office environments, e.g.,~\cite{wellner1994interacting, rekimoto1999augmented}). 
With the rise of affordable VR and AR HMDs these devices also have been explored 
as tools for assisting users when interacting with physical documents, e.g.,~\cite{grasset2007mixed, li2019holodoc}. 
Further, Grubert et al.~\cite{grubert2018office, ofektowards} and McGill et al.~\cite{mcgill2019challenges} explored the positive and negative qualities that VR introduces in mobile knowledge work scenarios. Desktop-based environments have been studied for tasks such as text entry~\cite{mcgill2015dose, knierim2018physical, grubert2018text},
system control~\cite{zielasko2019passive, zielasko2019menus} and visual analytics~\cite{wagner2018virtualdesk, buschel2018interaction}. Research on productivity-oriented desktop-based VR has concentrated on the use of physical keyboards and mouse-based input along with HMDs~\cite{schneider2019reconviguration,wang2020towards},
controllers and hands~\cite{kry2008handnavigator, zielasko2019passive}, and, recently, tablets~\cite{Biener2020Breaking, Gesslein2020Pen, surale2019tabletinvr}. 

Our work complements these prior studies by investigating the potential of editing slide-based presentations in VR using mobile devices such as tablets.






 
 


\subsection{Pen-based, In-air and Gaze-based Interaction}
Besides the commonly used single-point input with pens, enhanced interaction techniques have been explored. 
Examples include using touch input on the non-dominant hand, supporting pen input in bimanual interaction~\cite{hinckley2010pen+, pfeuffer2017thumb+},
unimodal surface-based pen-postures~\cite{cami2018unimanual}, bending~\cite{fellion2017flexstylus} or using sensors in or around the pen~\cite{hinckley2013motion, matulic2020pensight}
for gestures and postures, and examining pen-grips~\cite{hyong2011grips}.
Our work was inspired by tilting~\cite{tian2008tilt} and hovering~\cite{grossmann2006hover} the pen above interactive surfaces, which we use in a VR context. 


The use of pens in AR and VR has also been investigated as a standard input device on physical props~\cite{szalavari1997personal}, 
as well as using grip-specific gestures for in-air interaction~\cite{li2020grip}. 
The accuracy of pen-based in-air pointing has also been studied~\cite{pham2019pen}.
In AR, pen-based interaction was  specifically investigated for an object manipulation task~\cite{wacker2019arpen}.

In prior work on combining in-air with touch interaction, Marquardt et al.~\cite{marquardt2011continuous} investigated the use of on and above surface input on a tabletop. Chen et al.~\cite{chen2014air+} explored in-air use of on and above surface input on a tabletop. They propose that interactions can be composed by interweaving in-air gestures before, between, and after touch on a prototype smartphone augmented with hover sensing. Hilliges et al.~\cite{hilliges2009interactions} have been using hover to allow more intuitive interaction with virtual objects that represent physical objects. More recently, Hinckley et al.~\cite{hinckley2016pre-touch} have been exploring a pre-touch modality on a smartphone including the approach to record trajectories of fingers to distinguish between different operations. Such technology can be used to connect 3D tracking and touchscreen digitizer for better accuracy of tracking.

Most VR in-air interaction typically aims at using unsupported hands. To enable reliable selection, targets are designed to be sufficiently large and spaced apart~\cite{speicher2018VRselection}. Our focus on mobile knowledge workers on the move implies small gestures to reduce working fatigue and to retain usability in potentially cramped environments, such as airplane seats or tiny work places such as touchdown spaces. We utilize gestures to be used by a hand, resting on the screen of a tablet and holding a pen. 

In addition, the combination of eye-gaze with other modalities such as touch~\cite{pfeuffer2015gaze}, 
in-air gestures~\cite{pfeuffer2017gaze+, schweigert2019eyepointing} 
and head-movements~\cite{ kyto2018pinpointing, sidenmark2020bimodalgaze} 
has been investigated for interaction in spatial user interfaces. For a recent survey on gaze-based interaction in AR and VR, see Hirzle et al.~\cite{hirzle2019design}.

Specifically, our techniques were inspired by gaze-based interaction with virtual screens~\cite{Biener2020Breaking} as well as the combination of pen-based and touch-based interaction for mode switching~\cite{pfeuffer2017thumb+} but adapted those techniques specifically for the use case of editing presentations.

\subsection{Presenting and Authoring in XR}

XR has been explored for complementing or substituting established methods for presenting materials, e.g., in the medical domain~\cite{pantelidis2018virtual}, 
general education~\cite{kaminska2019virtual}
or training~\cite{checa2020review}.
For example, Kockro et al.~\cite{kockro2015stereoscopic} compared VR and (2D) PowerPoint lectures in an anatomy context, and found no performance differences but found VR to be rated higher in domains such as spatial understanding and enjoyability. With the recent rise of online conferences, VR has also been explored for delivering oral presentations and poster sessions~\cite{le2020enhancing}.
However, the benefits and drawbacks of presenting in VR compared to 2D conferencing tools such as Zoom, are yet to be explored in depth. 
XR has also been proposed as an aid for training public speaking~\cite{slater2006experimental} as well as in-situ support~\cite{parmar2020making}.

Besides using XR for presenting content to an audience, considerable work was invested in creating content for and in XR~\cite{macintyre2004dart, ashtari2020creating, nebeling2020xrdirector}.
Specifically, XR was investigated for supporting modelling~\cite{deering1995holosketch, reipschlager2019designar}, sketching~\cite{drey2020vrsketchin} 
and creating animations~\cite{arora2019magicalhands, cannavo2019immersive, vogel2018animationvr}. 

Complementary to these previous approaches, our work focuses on utilizing VR as a tool for authoring 2D presentations.

\section{Interaction Techniques for Authoring Presentations in Mobile VR}
\label{sec:Concept}
We looked at several challenging aspects of using a 2D slide editing program, such as dealing with 3D orientation and ordering, dealing with temporal data, and retrieving information from a large corpus of graphics data.
Then, we designed a set of interaction and visualization techniques using the advantages that VR provides, such as a large display space, a depth display and in-air interaction. Those techniques are just sample points in the entire space of tasks used for presentation authoring, but they can already show the advantages of using a VR environment. 
With our techniques, we want to support knowledge workers on the go or other confined spaces, limiting the choices of hardware. Therefore, our setup includes a tablet lying in front of the users, who hold a stylus with two buttons in their dominant hand, while their non-dominant hand is used for mode switches on the touchscreen in the area near the border. Both tablet and stylus are spatially tracked to represent them in VR.
For designing these interaction techniques, we followed an iterative approach with multiple design iterations consisting of conceptualization, implementation and initial user tests (eating your own dog food) \cite{unger2012project}.

As the techniques are designed to support users when working in adverse conditions, such as confined spaces, lack of privacy and limited display size~\cite{grubert2018office, Biener2020Breaking}, VR provides multiple advantages. First, users are no longer restricted to their available physical displays and can view information in the space around them, beyond the bounds of a mobile device. Second, the three-dimensional VR display enables depth visualization to allow utilization of the space above (or below) a 2D surface. Third, touch-based interaction can be complemented with further modalities such as in-air or gaze-based interaction.  Fourth, the entire display is seen only by the users, maintaining their privacy and reducing visual disturbance from the environment.

Because we want to support small (mobile) work spaces, we use the tablet's surface as the main interaction space. This provides space for hand motions above it while the touchscreen supports easy and accurate input.
However, in many cases it might not be comfortable to look at the tablet lying on a table. In such cases, the slide and also the pen could be re-projected to be in front of the user's head for indirect manipulations as suggested by prior work~\cite{grubert2018text}.

In the following subsections, we will present concepts for editing an individual slide, as well as concepts for working with multiple slides and other resources.
Note: For video description of interaction techniques refer to the accompanying video.

\subsection{Editing Slides}

Preparing presentations requires the user to create slides that contain a collection of information, arranged both along the area of the slide, as well as in depth (layering of items) and in time (animation of the items). In the following, we propose techniques for such tasks using a pen and a tablet in VR.



\subsubsection{Manipulation of Objects}
\label{sec:ManipulationObjects}
Common presentation tools, such as Microsoft PowerPoint, let the user create slides that may contain a collection of items such as text, images, videos, three-dimensional objects and more. 
Such items can be selected, and dragged to change their position on the slide. They can be rotated and scaled using dedicated widgets and may even be rotated in three-dimensions using additional input fields or symbolic input.

We explore how to use a pen alongside touch input in VR to provide a unifying interface for 2D and 3D object manipulation of elements on the slide.
We propose to use a pen that is spatially tracked which expands the interaction space to include not only the tablets surface but also the space around the user. 
This can potentially make object manipulations more intuitive.

\begin{figure}[t]
	\centering 
	\includegraphics[width=1\columnwidth]{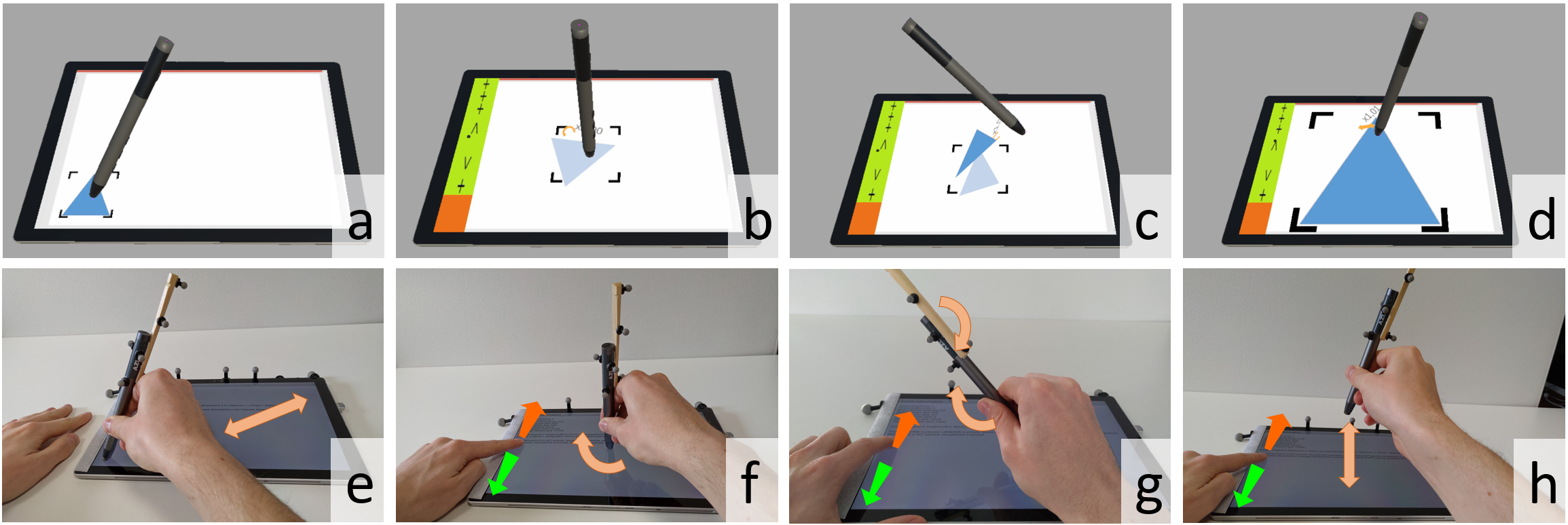}
	\caption{
	Tracking of the stylus' six degrees of freedom in VR enables complex manipulation of an object on the slide. The dominant hand holding the stylus is used to drag, rotate, scale and 3D rotate the object, while the non-dominant hand's touch on the screen's bezel-area (green area on the left) modulates the interaction. Without any bezel touch (a, e) objects are dragged by the stylus. Bezel touch and rotation of the stylus along its axis while it touches the screen (b, f) rotates the object on the screen's plane. Rotating the stylus in the air while keeping the bezel touch (c, g) rotates the objects in 3D. Finally, touching the bezel while the stylus is in the air and moving the stylus towards or away from the screen (d, h) decreases or increases the scale of the object uniformly.
    }
	\label{fig:ObjectManipulation}
	\vspace{-0.3cm}
\end{figure}

Translating a selected object is supported by standard drag and drop using a stylus or the user's finger.
An object can be rotated by rotating the pen (as if it is attached to the pen) and it can be scaled by moving the pen further away from or closer to the tablet (as if pulling to increase the size). 
To differentiate between translation, rotation and scaling, a finger of the non-dominant hand on the bezel of the touchscreen is used to control the modality of the manipulation, as shown in Figure \ref{fig:ObjectManipulation}. We chose a bezel-based technique, as they have already been successfully used for mode-switches in other scenarios (e.g. \cite{Biener2020Breaking, Gesslein2020Pen}). When no finger touches the designated area on the bezel of the screen, the stylus is used to select an object and to drag it to a new position (Figure \ref{fig:ObjectManipulation}, a and e). Touching the bezel with the non-dominant hand, while still touching an object with the stylus, activates the 2D rotation mode.
Rotating the stylus around its axis, while still touching the surface with the tip of the stylus, will rotate the object in the screen plane (Figure \ref{fig:ObjectManipulation}, b and f). 

Lifting the pen away from the surface, while still touching the bezel in rotation mode, enables the user to perform 3D rotations by rotating the stylus in space.
When performing a 3D rotation, two instances of the object are displayed - one as a flat projection on the slide, and a full 3D display of the object in the air above the screen, enabling the user to better see the three-dimensional pose (Figures \ref{fig:ObjectManipulation}, c and g and \ref{fig:teaser}, a). 
The position of the 3D display is fixed above the objects position on the slide and its rotation is determined by the rotation of the stylus.



To scale an object, the non-dominant hand touches the bezel, while the tip of the stylus is in the air not touching the screen. Moving the stylus' tip up, away from the screen, increases the scale of the object uniformly, while bringing the tip closer to the screen reduces the its scale (Figure \ref{fig:ObjectManipulation}, d and h).

The sensitivity of these manipulations can be controlled by moving the finger's vertical location on the bezel (see Figure \ref{fig:ObjectManipulation}, f, g and h). Moving the touch point up increases the control-display gain of the rotation and the scaling, enabling larger changes with small pen movements. Moving the finger down enables better accuracy.
With this form of object manipulation there is no need to select potentially small scale and rotation handles on already small display sizes because of the bimanual mode activation technique. For the case of 3D rotations in particular, the actual 3D preview of the rotation above the object combined with VR-based head movement and three-dimensional display is something not possible in classical 2D presentation tools and has the potential to make 3D rotations more intuitive for the user. 

\subsubsection{Occlusion Handling}
\label{sec:HandlingOcclusions}
Most presentation tools support explicit ordering of objects, where objects of a higher layer occlude objects of lower layers. As a result, objects may become partially or even completely hidden behind other objects, making visual identification difficult or impossible, and increasing selection difficulty. While some applications such as PowerPoint display 
a separate list of all objects on a slide, sorted by their levels, such a list requires the user to create a mental map, matching the location of each item on the list to the visual objects on the slide. An example can be seen in Figure \ref{fig:ReorderingTaskConditions}, c which shows a simplified version of the PowerPoint interface used in our study. 


The Mac version of PowerPoint uses 
a dynamic reordering mode\footnote{https://www.indezine.com/products/powerpoint/learn/shapes/dynamic-reorder-of-overlapping-shapes-in-ppt2011-mac.html Last access September 1, 2021} that displays objects with their layers slightly rotated, similar to Figure \ref{fig:ReorderingTaskConditions}, b.  
In this mode, the layers can be grabbed and moved to rearrange their order.
However, with increasing number of objects this can also become challenging as the layers can potentially overlap and make it harder to see which object belongs to which layer, impacting selection accuracy.

\begin{figure}[t]
	\centering 
	\includegraphics[width=0.65\columnwidth]{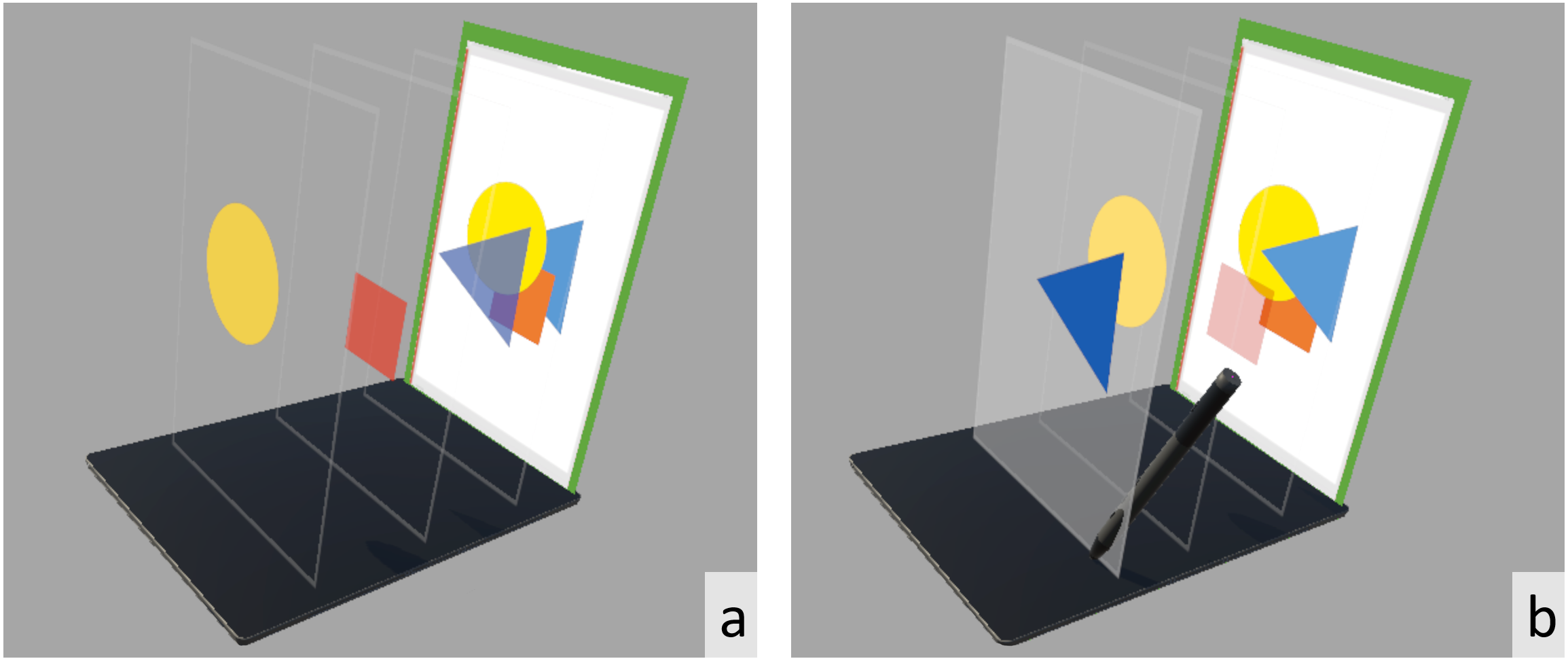}
	\caption{Reordering layered objects. The layers are displayed with the top one on the far left of the tablet, and the bottom one on the right. On the right end of the tablet, we display the full composited slide for reference. a) An example of side view with a slide containing a yellow circle on top, than a red square followed by a blue triangle. b) A user dragging the blue triangle to place it on the top layer.}
	\label{fig:handlingOcclusions}
	\vspace{-0.3cm}
\end{figure}

Inspired by this dynamic reordering mode, we propose to use the space above the tablet in VR to present the object-layers to the user in a way that facilitates assessing how the objects are ordered. 
Specifically, we propose to rotate the object-layers by 90 degrees, move them up to stand on top of the tablet and slightly separate them (see Figures \ref{fig:handlingOcclusions} and \ref{fig:teaser}, b). 
The tablet and the user's head can be repositioned in VR to resolve any potential occlusion issues that might arise from a fixed viewpoint. 
We also conducted informal experiments with further degrees of rotation. Zero degrees of rotation (layer parallel to the display) did not scale beyond a few layers due to the imprecise nature of in-air selection compared to touch-based selection. When comparing different amounts of rotations (0, 45 and 90 degrees), 90 degrees was perceived as most comfortable and efficient.  Furthermore, the selection process can be supported by showing the intersections of the layers with the touchscreen (lines in Figures \ref{fig:handlingOcclusions} and \ref{fig:teaser}, b). Thus, instead of in-air selection, layers can be selected by touching their intersection lines, and precisely dragged to a new position (Figure \ref{fig:handlingOcclusions}, b). Using touch on the tablet display allows for precise interaction, even when the number of objects increases 
and the layers get closer to each other.  
There is also a projection of the complete slide to the right of the object layers which presents the current layout in 2D and facilitates the understanding of the ordering.
In our implementation, the layer reordering mode is toggled by touching the lower left corner of the bezel area that is also used for mode switching when manipulating objects. An experiment showing superior performance of this technique compared to baseline Powerpoint implementations is presented in section \ref{sec:ReorderStudy}.

\subsubsection{Animations}
\label{sec:Animations}
Presentation applications such as PowerPoint show keyframes of object animations as locations on a slide. 
As the area of the slide is limited and may quickly be overloaded with information, PowerPoint displays only the start and end keyframes and only for the currently selected object. 
We facilitate visualizing and editing animations through a 3D visualization where the extra spatial dimension represents time. 

\begin{figure}[t]
	\centering 
	\includegraphics[width=0.65\columnwidth]{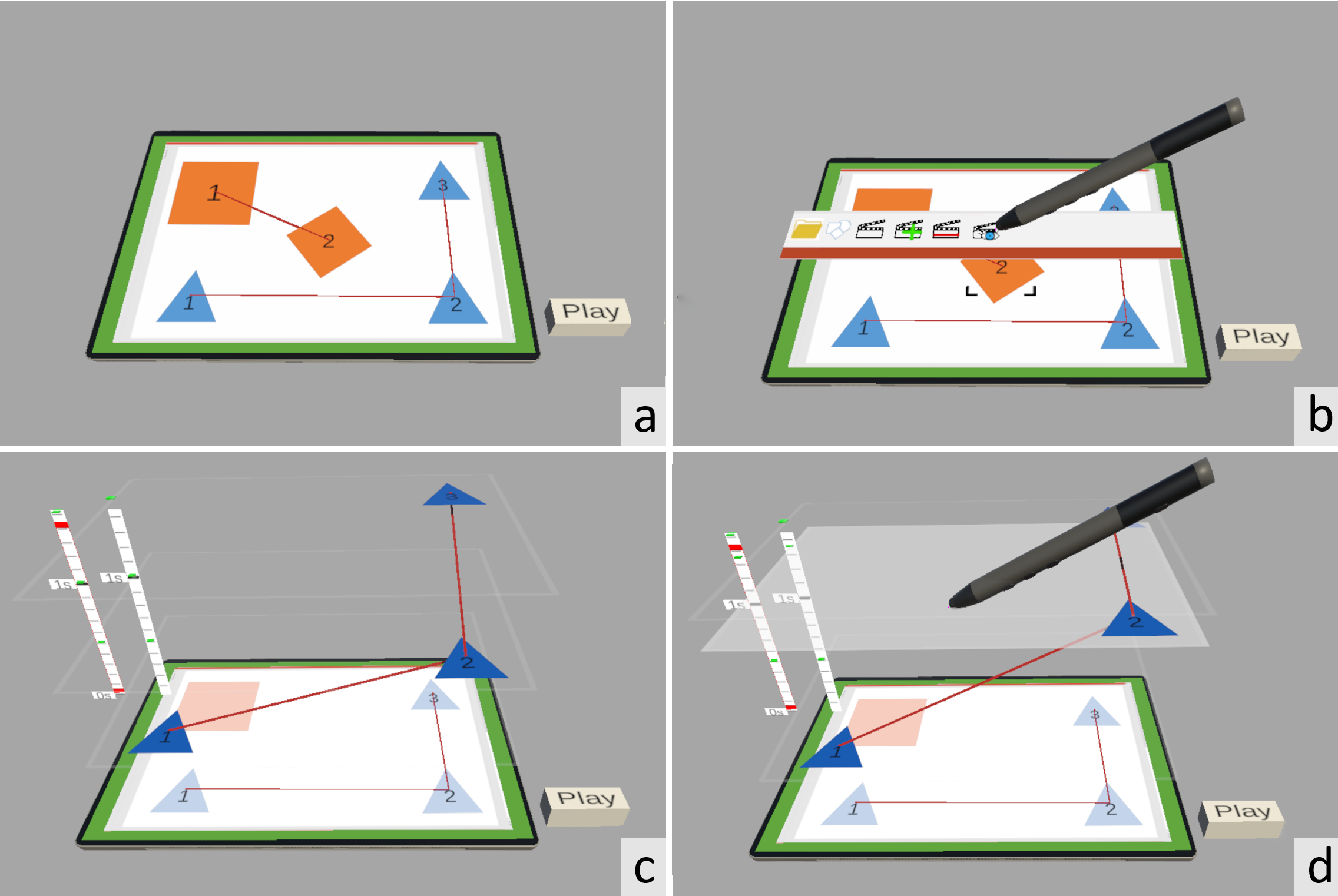}
	\caption{a) Animation mode showing the keyframes of two objects on the slide. b) Menu that can be used to add or delete keyframes or make keyframes invisible. c) 3D animation mode which shows the keyframes as layers with their position in z-direction representing time. d) Moving the second keyframe up so the movement from the first to the second will be slower.}
	\label{fig:animationConcept}
	\vspace{-0.3cm}
\end{figure}

By selecting an entry in an in-air menu opened by the second pen button, the user enters an explicit animation mode. As a base 2D visualization, we employ a similar approach to PowerPoint, visible in Figure \ref{fig:animationConcept}, a, but showing all keyframes of all objects by default, instead of showing it only for the selected objects as in PowerPoint. However, the keyframes of certain objects can be hidden via a menu entry, which appears next to the pen after pressing the second pen button, as can be seen in Figure \ref{fig:animationConcept}, b. The keyframes may be manipulated (dragged, rotated and scaled) like any other item on the slide. To represent time, the keyframes are numbered and connected by lines representing the sequence of the animation (Figure \ref{fig:animationConcept}).

By using the space above the screen, we can also visualize the time of each keyframe by locating it at the corresponding height above the screen. The further up the keyframe is placed, the later the associated object state will be reached in the animation. This enables us to display all objects and the relations between their animations. All keyframes of an animation are connected with a colored curve that shows the path the animation takes in time (height), as visible in Figures \ref{fig:animationConcept}, c and \ref{fig:teaser}, c.

Similar to our reordering technique, the three-dimensional time view of the animation is toggled by touching the lower-left corner of the tablet's bezel area while the animation mode is active. A keyframe can be grabbed in-air using the first button on the stylus and moved up and down to change the corresponding time (Figure \ref{fig:animationConcept}, d). 
For the animation interface, non-rotated layers were used (in contrast to occlusion handling), because object manipulation on the screen is possible while the 3D time view is active, overloading the touch inputs that would otherwise need to be used to move rotated layers. Also, animating depends, in contrast to pure reordering, much more on the actual position of the objects on the slide. Therefore, we prioritized spatial consistency over ease of selection.

Vertical timelines, displayed left of the tablet, help to indicate a keyframe's precise time. Applying a two finger pinch or move gesture in the bezel area can scale or scroll through the displayed animation along the time axis to better see parts of the animation. We display two timelines: one that displays the global animation timeline, and the other displays the currently viewed section of the timeline (Figure \ref{fig:animationConcept}, c and d). In addition to toggling visibility of animations, the menu opened by the second pen button is also used to add and remove keyframes (Figure \ref{fig:animationConcept}, b). A play button that can be touched by the stylus can play or pause the animation. The default display of the animation is on the tablet screen, but it is also possible to render it on a virtual screen placed away from the tablet for easier viewing.

\subsection{Working Across Slides}
While the tablet screen is the default surface to display and edit the current slide, authoring presentations requires to also access further data sources, e.g., the slides of the currently edited slide-set, fetching resources from other presentations, or browsing external content such as images or PDF files. In the following, we discuss means for viewing further media concurrently to the active slide, and transferring content between these displays and the active slide screen.

\subsubsection{Slide Overview}
\label{sec:slideoverview}
When authoring presentations, it is a common task to copy content between slides or to go back and forth between slides to review the content.
Especially on a small display in presentations with several slides, it is often not possible to fit all slides on the screen with sizes that allow the user to recognize them, and the user has to scroll through the slide overview (or slide sorter view) to find the target slide. The large display space in VR can be used to mitigate this problem. We use two different ways to access other slides on the slideshow. First, while the current slide is displayed on the tablet screen, slides before and after the current slide are displayed to the left and right of the tablet, see Figure \ref{fig:teaser}, d. This technique is inspired by similar visualizations of Gesslein et al. \cite{Gesslein2020Pen}. Second, a slide overview of the current slide-set is displayed in front of the user, as can also be seen in Figure \ref{fig:teaser}, d. This overview can be scrolled and zoomed using a two-finger swipe or pinch gesture in the border area of the tablet while gazing at the overview area. While it would be possible to select a slide from the overview using in-air gestures, they may be exhaustive and may not fit limited work spaces. Instead, we use eye gaze to pre-select a slide of interest, indicated by a green frame, and while maintaining the gaze, swipe down on the touchscreen to confirm selection. The selected slide is then displayed as the active slide in the current slide set.

\subsubsection{Multiple Content Sources}
\label{sec:MultipleSlideSets}
When creating a presentation, it is very common to use additional content
such as text, images or videos. Just like displaying the slide overview, the space in front of the user can be used to display a myriad of content like images, videos or web pages that can be added to the presentation. The user can add such content areas through a menu opened, again, by a the second pen button. It is also possible to fade out some of these areas while they are not used by clicking on the corresponding toggle button with the pen (red buttons in front of the tablet in Figure \ref{fig:teaser}, d).

\subsubsection{Copying Content}
\label{sec:ContentTransfer}

\begin{figure}[t]
	\centering 
	\includegraphics[width=1\columnwidth]{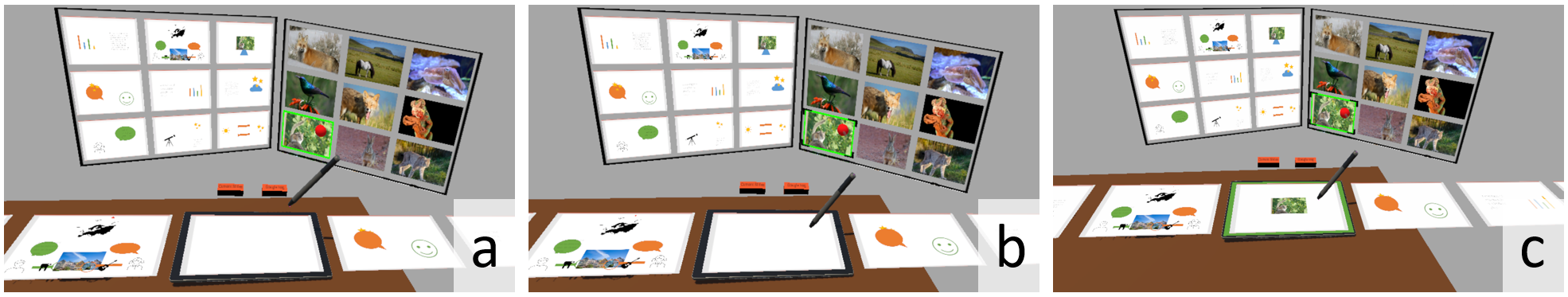}
	\caption{a) The user is looking at an object (image) and sees the position of the pen above the tablet as a red dot. b) The user touches the tablet with the pen to select the image. c) The user looks back down and the selected object is copied to the currently edited slide.}
	\label{fig:ContentTransferSlideOverview}
\end{figure}

The user can select objects from the external content displays described in Section \ref{sec:MultipleSlideSets} and copy them to the currently active slide. As the user gazes at an element such as a slide, a semi-transparent copy of the pen is visualized in the same relative pose to the gazed element as the physical pen is located above the physical tablet, and the touch location which it is hovering over is shown as a red dot (Figure \ref{fig:ContentTransferSlideOverview}, a). As the user moves the stylus above the tablet screen, the corresponding copy above the gazed element moves as well. 
The user can select an object from content displays in the same way as selecting an object on the tablet -- by touching it as visualized in Figure \ref{fig:ContentTransferSlideOverview}, b. Looking back at the tablet while still touching the screen with the stylus will copy the selected object to the current slide (Figure \ref{fig:ContentTransferSlideOverview}, c).
Similarly, the user can copy an entire slide from the current or another slide set by touching this slide at a position where no object is placed. Looking down will insert this slide after the currently edited slide. Both objects and slides can also be copied through the menu opened by the second pen button.


\section{Performance Evaluation}
\label{sec:PerformanceEvaluation}
The two main advantages of VR for knowledge worker tasks that we focus on in this paper are the larger display space and the three-dimensional  viewing of 3D content.
To evaluate their advantages for presentation authoring, we chose two tasks, each representing one of the above mentioned advantages: using the large field of view for displaying an overview of content and using the 3D space above the tablet for ordering slide objects that occlude each other.   
These two tasks represent issues that current users are often faced with: searching for a slide on a limited screen and ordering occluded objects. However, the presented concepts can also be generalized to other tasks.
The evaluation was done by conducting two separate lab-based studies using a within-subjects design in each case.

\begin{figure}[t]
	\centering 
	\includegraphics[width=1\columnwidth]{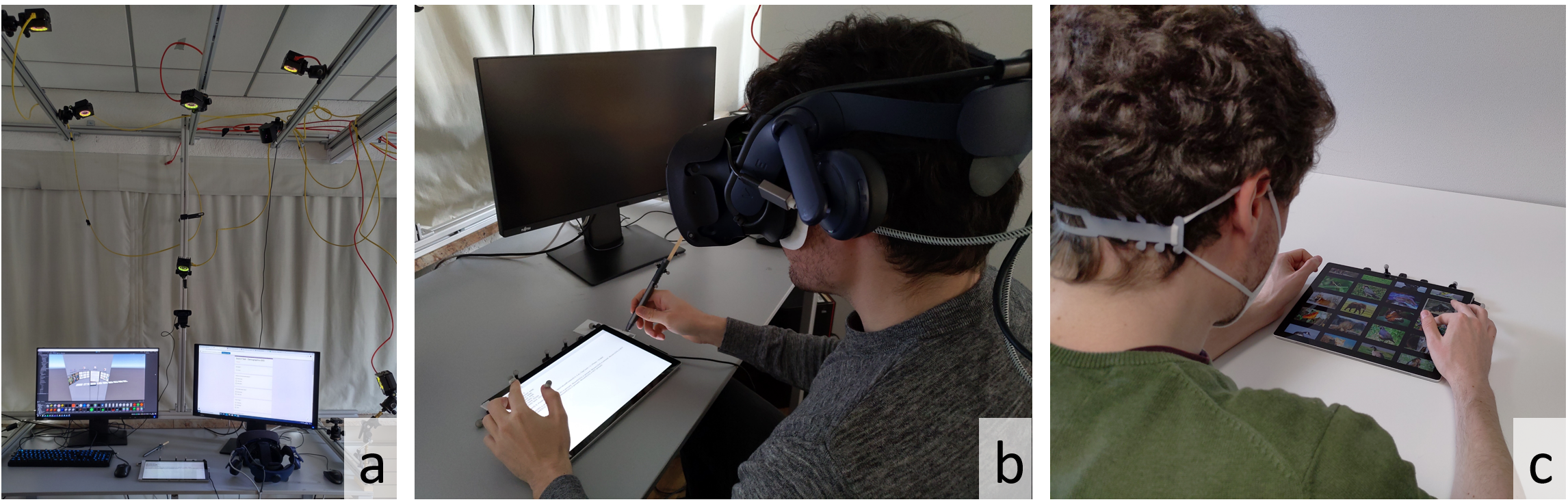}
	\caption{a) Study setup with 6 OptiTrack cameras and two VIVE lighthouse basestations (one is behind). OptiTrack cameras were used for the reordering task and the usability evaluation, while lighthouse basestations were used for the search task. b) Participant in the reordering task and the usability evaluation. c) Participant in the tablet search task condition. }
	\label{fig:studySetup}
	\vspace{-0.3cm}
\end{figure}

\subsection{Search Study}
Virtual Reality HMDs enable the user to have a large display space around them, which can be used to show a large number of objects, such as slides in a slide sorter view or images during an image search, at the same time as shown in section \ref{sec:MultipleSlideSets}. In contrast, using a small tablet screen may force the user to scroll to be able to view a similar amount of items with a comparative size. We selected a search task to quantify the possible advantages of VR HMDs compared to a tablet screen. The participants were presented with a target image and then had to find and select it among 63 other images. We chose a corpus of images of different animals. While this kind of visual stimuli are commonly used assets in presentations, they represent challenging content due to the amount of details in naturalistic images.

\begin{figure}[t]
	\centering 
	\includegraphics[width=1\columnwidth]{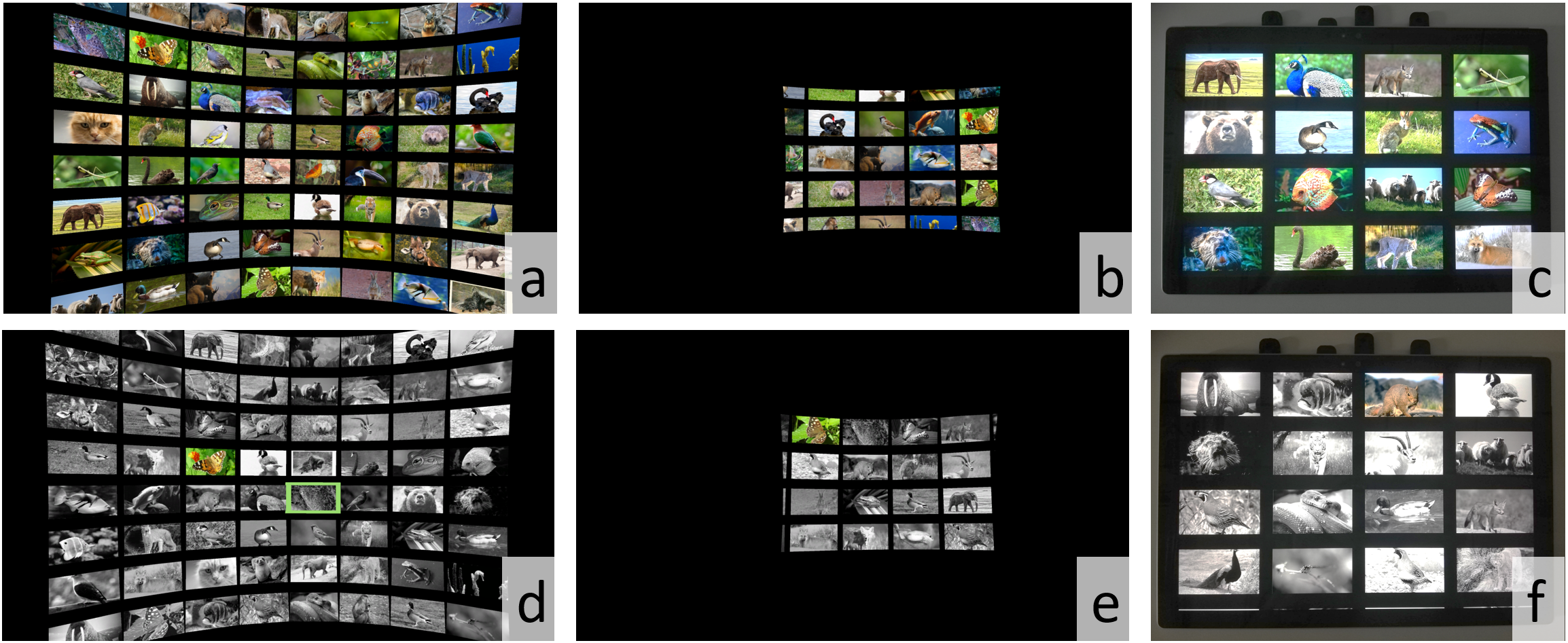}
	\caption{Conditions for the visual search task: \textsc{hard} conditions, with colored images only with a) \textsc{vr-full}, b) \textsc{vr-limited} and c) \textsc{tablet} \textsc{interface}; \textsc{easy} conditions with only the target image colored with d) \textsc{vr-full} e) \textsc{vr-limited} and f) \textsc{tablet} \textsc{interface}. The green frame in d) indicates the image that is currently selected via gaze.}
	\label{fig:SearchTaskConditions}
	\vspace{-0.3cm}
\end{figure}

The experiment consisted of two independent variables: \textsc{interface} and \textsc{difficulty}.
We used three types of \textsc{interface}s. The first, \textsc{vr-full} used the maximum field of view (FoV) provided by the HMD. The second, \textsc{vr-limited}, artificially limited the users' FoV in VR to reflect a similar FoV to that of the tablet's display. And the third was \textsc{tablet}, using the actual tablet display without VR.
The second independent variable was \textsc{difficulty}. The levels were \textsc{hard}, using the original colored images, leading to an attentive search, and \textsc{easy}, where all images but the target image were reduced to grayscale making the search in this condition pre-attentive, so the target could be immediately spotted~\cite{wolfe2017five}. The combination of these two variables leads to six conditions which are depicted in Figure \ref{fig:SearchTaskConditions}. The dependent variables measured in this experiment were task completion time, number of errors, usability (System Usability Scale, SUS)~\cite{brooke1996sus}, workload (NASA TLX unweighted version)~\cite{hart1988development}, and simulator sickness (SSQ)~\cite{kennedy1993simulator}.

\subsubsection{Participants}
Twenty participants took part in this study (6 female, 14 male), with a mean age of 28.85  ($SD=5.2$). Five participants wore glasses during the experiment and all but one had at least some prior VR experience.

\subsubsection{Apparatus}
In all conditions the participants were presented with a set of images. Each image was set to occupy approximately 8 degrees horizontally of the participant's field of view, both displayed on the tablet's screen or HMD. While it is known that image size impacts image search~\cite{hurst2011size}, we empirically determined this size to be both legible and selectable in VR and on the tablet. With an approximate distance to the tablet of 40~cm this resulted in an image width of 5.8~cm on the tablet. Sixty-four images were arranged in four columns, such that 16 images were fully visible without a need for scrolling. 

In both VR conditions (\textsc{vr-full} and \textsc{vr-limited}) the images were placed on a sphere around the user at a distance of 75 cm which resembles the focal distance of the HTC Vive HMD, resulting in an image width of 10 cm.
In the VR conditions, the images were arranged in 8 columns covering 65 degrees horizontally and 45 degrees vertically to enable participants to comfortably reach all images with eye-gaze and only slight head movements. The field of view of the \textsc{vr-limited} condition was artificially limited using black planes in all four directions, resulting in a field of view of about 36 x 24 degrees, resembling the field of view of the tablet (26 x 17.5~cm at a distance of 40~cm). Both the tablet and the VR applications were implemented using Unity 2019.4. 
The \textsc{tablet} conditions were performed on a Microsoft Surface Pro 4 as shown in Figure \ref{fig:studySetup}, c. For the VR conditions, we used a HTC Vive Pro Eye, which provides built-in eye-tracking and two lighthouse base stations (Figure \ref{fig:studySetup}, a). We combined it with the Microsoft Surface to enable touch input. In contrast, to the further studies, the Optitrack system depicted in \ref{fig:studySetup}, a, was not used in this study.

\subsubsection{Procedure}
The study started by asking the participants to sign a consent form and fill out a demographic questionnaire.
The order of the six conditions was balanced using a balanced Latin square.
In each condition the participants were first presented with the target image, either on the tablet screen in the \textsc{tablet} conditions or in the air in front of the user in VR. Upon touching the tablet's screen the target image vanished and the image search started.
In the \textsc{tablet} condition, the participants had to scroll through the images by touching and dragging on the display. When they found the target, they selected it by touching it. In the VR (\textsc{vr-full} and \textsc{vr-limited}) conditions, the participants had to search for the target by moving their eyes and head and select the target by eye-gaze and confirm the selection by taping anywhere on the tablet. Prior to the start of the VR conditions, users conducted eye-gaze calibration using the built-in calibration routine of the HTC Vive Pro Eye.
After selecting an image, the next target image was displayed. In each condition, participants had to find 30 images, which were always the same but in randomized order and positions while ensuring that targets are positioned in all regions. After completing a condition, participants answered the simulator sickness questionnaire~\cite{kennedy1993simulator}, the system usability scale questionnaire~\cite{brooke1996sus} and the NASA task load index~\cite{hart1988development}. Also, we recorded the task completion times and errors for each task. On average, it took 45 minutes to complete this experiment.

\subsubsection{Results}
Repeated measures analysis of variance (RM-ANOVA) was used to analyze task completion times, which were non-normal and therefore log transformed. For multiple comparisons Bonferroni adjustments were used at an initial significance level of $\alpha=0.05$. Aligned Rank Transform~\cite{wobbrock2011aligned} was used for subjective data and errors that are not normally distributed (or could not be normalized using log transform). The main results are displayed in Table \ref{tab:resultsTableSearch}.

For each participant an average task completion was computed from the 30 tasks, including trials with errors.
Analyzing the task completion times indicated significant simple main effects of \textsc{interface} and \textsc{difficulty} on task completion time and that there were also significant interaction effects. Specifically, the \textsc{vr-full} ($M=4.7s$, $SD=3.52$) conditions were significantly faster than the \textsc{vr-limited} ($M=5.47s$, $SD=3.34$) and \textsc{tablet} ($M=5.75s$, $SD=2.56$) conditions. 
Also, a significant difference could be found between \textsc{vr-limited} and \textsc{tablet}. As expected, the \textsc{easy} ($M=2.65s$, $SD=1.02$) conditions were significantly faster than the \textsc{hard} ($M=7.97s$, $SD=2.2$) conditions.

\begin{figure}[t]
	\centering 
	\includegraphics[width=0.55\columnwidth]{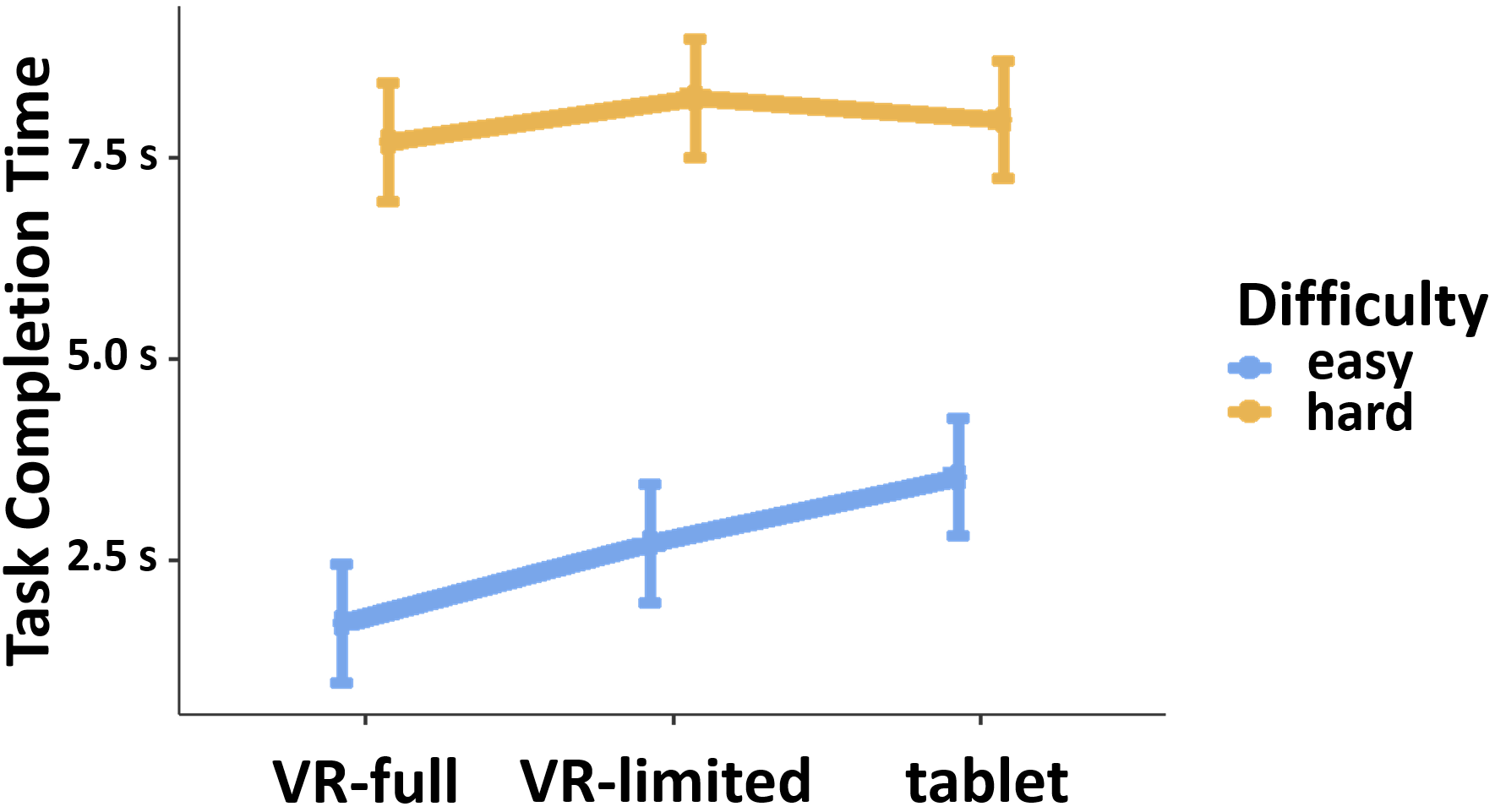}
	\caption{Task completion times for the six conditions, which is significantly influenced by the interface in the easy conditions, but not in the hard.}
	\label{fig:searchTaskInteractionEffect}
	\vspace{-0.3cm}
\end{figure}

The interaction effect is visible in Figure \ref{fig:searchTaskInteractionEffect}. Post-hoc comparisons showed that for the \textsc{easy} conditions, \textsc{vr-full} ($M=1.71s$, $SD=0.61$) is significantly faster than \textsc{vr-limited} ($M=2.7s$, $SD=0.8$) and both VR methods are significantly faster than \textsc{tablet} ($M=3.53s$, $SD=0.66$).  This suggests that the wider FoV makes the search faster, as indicated by prior work~\cite{ragan2015effects}. The different input techniques  (gaze vs. scrolling) are likely to contribute to difference between the \textsc{vr-limited} and \textsc{tablet}. However, for the \textsc{hard} conditions, no significant differences between the \textsc{interfaces} were indicated. This could suggest that the larger FoV does not necessarily provide an advantage in a search were all possible targets have to be looked at in detail.

\textsc{interface} also significantly influenced the number of errors. Specifically, participants made significantly less errors in the \textsc{tablet} ($M=0.8$, $SD=1.11$) conditions compared to both the \textsc{vr-full} ($M=5.08$, $SD=4.65$) and \textsc{vr-limited} ($M=3.58$, $SD=3.64$) conditions. 
There was no significant difference between \textsc{vr-limited} and \textsc{vr-full}. Also, the analysis showed that \textsc{difficulty} had a significant effect on the number of errors in such a way that the \textsc{easy} conditions ($M=2.63$, $SD=3.81$) resulted in significantly less errors than the \textsc{hard} conditions ($M=3.67$, $SD=3.9$). However, the error rate in all conditions was rather low and higher error rates in the VR conditions could be explained by the eye-gaze technique, relying on off-the shelf gaze-tracking in the HTC Vive Pro Eye. It is not always perfectly accurate and if the user is not fully concentrated, the gaze might lose the target in the moment of the confirmation, resulting in a wrong selection. However, there are possibilities for improvement like a delay before switching the selected images or a dwell time. More errors in the \textsc{hard} conditions can be explained by user mistakes, for example when looking for an orange fish they chose the yellow one.

Analyzing the results from the total severity dimension of the simulator sickness questionnaire indicated a significant influence of \textsc{interface} and \textsc{difficulty}. However, pairwise comparisons showed no significant differences. Unsurprisingly, a significant influence of \textsc{difficulty} on the usability score was detected in such a way that the \textsc{hard} ($M=85.88$, $SD=13.07$) conditions had a significantly lower usability than the \textsc{easy} ($M=89.75$, $SD=10.69$) conditions. However, no significant influence of \textsc{interface} could be detected, regarding usability. The \textsc{difficulty} also had a significant influence on the overall task load, such that the task load was significantly higher for the \textsc{hard} conditions ($M=29.65$, $SD=16.54$) than for the \textsc{easy} conditions ($M=17.81$, $SD=12.73$). This makes sense, because compared to the \textsc{hard} conditions, in the \textsc{easy} conditions less mental effort is required to find the target. Again, no significant influence of \textsc{interface} could be detected.

\begin{table}[t]
    \centering 
    \tiny
    \begingroup
    \setlength{\tabcolsep}{5pt}
        \begin{tabular}{|>{\centering}m{0.45cm}||>{\centering}m{0.3cm}|>{\centering}m{0.3cm}|c|c|c||>{\centering}m{0.3cm}|>{\centering}m{0.3cm}|c|c|c|}
            \multicolumn{11}{c}{\small\bfseries\textbf{Search Task - Subjective Ratings}} \\
            \hline 
            & \multicolumn{5}{c||}{TS-SS} & \multicolumn{5}{c|}{SUS} \\
            \hline 
            & d$f_{1}$ & d$f_{2}$ & F & p &  $\eta^2_p$ & d$f_{1}$ & d$f_{2}$ & F & p &  $\eta^2_p$ \\ 
            \hline 

            I & \cellcolor{lightgray}$2$ & \cellcolor{lightgray}$38$ & \cellcolor{lightgray}$3.59$ & \cellcolor{lightgray}$0.04$ & \cellcolor{lightgray}$0.16$ & $2$ & $38$ & $0.24$ & $0.79$ & $0.01$   \\ 

            D &  \cellcolor{lightgray}$1$ & \cellcolor{lightgray}$19$ & \cellcolor{lightgray}$7.23$ & \cellcolor{lightgray}$0.01$ & \cellcolor{lightgray}$0.28$ & \cellcolor{lightgray}$1$ & \cellcolor{lightgray}$19$ & \cellcolor{lightgray}$8.25$ & \cellcolor{lightgray}$0.01$ & \cellcolor{lightgray}$0.3$ \\ 

            \hline 

            \tiny{I $\times$ D} &  $2$  & $38$  & $2.43$  & $0.1$  & $0.11$    &  $2$  & $38$  & $0.7$  & $0.5$  & $0.04$   \\ 

            \hline 
        \end{tabular}

        \begin{tabular}{|>{\centering}m{0.45cm}||>{\centering}m{0.3cm}|>{\centering}m{0.3cm}|c|c|c|}
            \hline 
            &  \multicolumn{5}{c|}{Overall task load}\\
            \hline 
            & d$f_{1}$ & d$f_{2}$ & F & p &  $\eta^2_p$  \\ 
            \hline 

            I &  $2$ & $38$ & $0.42$ & $0.66$ & $0.02$  \\ 

            D &   \cellcolor{lightgray}$1$ & \cellcolor{lightgray}$19$ & \cellcolor{lightgray}$40.51$  & \cellcolor{lightgray}$<0.001$  & \cellcolor{lightgray}$0.68$ \\ 

            \hline 

            \tiny{I $\times$ D} &  $2$   & $38$  & $1.98$  & $0.15$ & $0.09$ \\ 

            \hline 
        \end{tabular}

        \begin{tabular}{|>{\centering}m{0.45cm}||>{\centering}m{0.23cm}|>{\centering}m{0.3cm}|c|c|c||>{\centering}m{0.23cm}|>{\centering}m{0.3cm}|c|c|c|}
            \multicolumn{11}{c}{\small\bfseries\textbf{Search Task - Performance Data}} \\
            \hline 
            &\multicolumn{5}{|c||}{Task Completion Time} & \multicolumn{5}{c|}{Errors} \\
            \hline 
            & d$f_{1}$ & d$f_{2}$ & F & p &  $\eta^2_p$ & d$f_{1}$ & d$f_{2}$ & F & p &  $\eta^2_p$\\ 
            \hline 

            I & \cellcolor{lightgray}$2$  & \cellcolor{lightgray}$38$ & \cellcolor{lightgray}$32.3$ & \cellcolor{lightgray}$<.001$  & \cellcolor{lightgray}$.63$ & \cellcolor{lightgray}$2$ & \cellcolor{lightgray}$38$ & \cellcolor{lightgray}$33.96$ & \cellcolor{lightgray}$<.001$ & \cellcolor{lightgray}$.64$   \\ 

            D & \cellcolor{lightgray}$1$ & \cellcolor{lightgray}$19$ & \cellcolor{lightgray}$782.5$ & \cellcolor{lightgray}$<.001$ & \cellcolor{lightgray}$.98$  & \cellcolor{lightgray}$1$ & \cellcolor{lightgray}$19$ & \cellcolor{lightgray}$22.27$ & \cellcolor{lightgray}$<.001$& \cellcolor{lightgray}$.54$   \\ 

            \hline 

            \tiny{I $\times$ D} & \cellcolor{lightgray}$2$  & \cellcolor{lightgray}$38$ & \cellcolor{lightgray}$33.3$  & \cellcolor{lightgray}$<.001$  & \cellcolor{lightgray}$.64$ & $2$  & $38$   & $2.74$  & $.08$  & $.13$     \\ 

            \hline 

        \end{tabular}

    \endgroup

    \caption{RM-ANOVA results for the search task.  Gray rows show significant findings. I = \textsc{Interface}, D = \textsc{Difficulty}. TS-SS: Total Severity Dimension of the Simulator Sickness Questionnaire. SUS: System Usability Scale. d$f_1$ = d$f_{effect}$ and d$f_2$ = d$f_{error}$.}
    \label{tab:resultsTableSearch}
    \vspace{-0.3cm}
\end{table}

\subsection{Reordering Study}
\label{sec:ReorderStudy}
To quantify the benefits of a three-dimensional visualization of layered information, we used the reordering task to compare the standard \textsc{PowerPoint} tool for reordering object layers with the \textsc{dynamic reordering} tool implemented in PowerPoint for Mac and our \textsc{VR} tool. 
The participants were presented with a number of objects on a slide.
The displayed objects always included one yellow square and one red circle while the remaining objects were blue triangles.
The task for the participants was to bring the red circle directly in front of the yellow square, by moving the corresponding layers to the front or to the back.  
To reproduce many different scenarios, the objects were placed with different amounts of overlap, the number of layers was varied and the target object was placed at different depths.

\begin{figure}[t]
	\centering 
	\includegraphics[width=1\columnwidth]{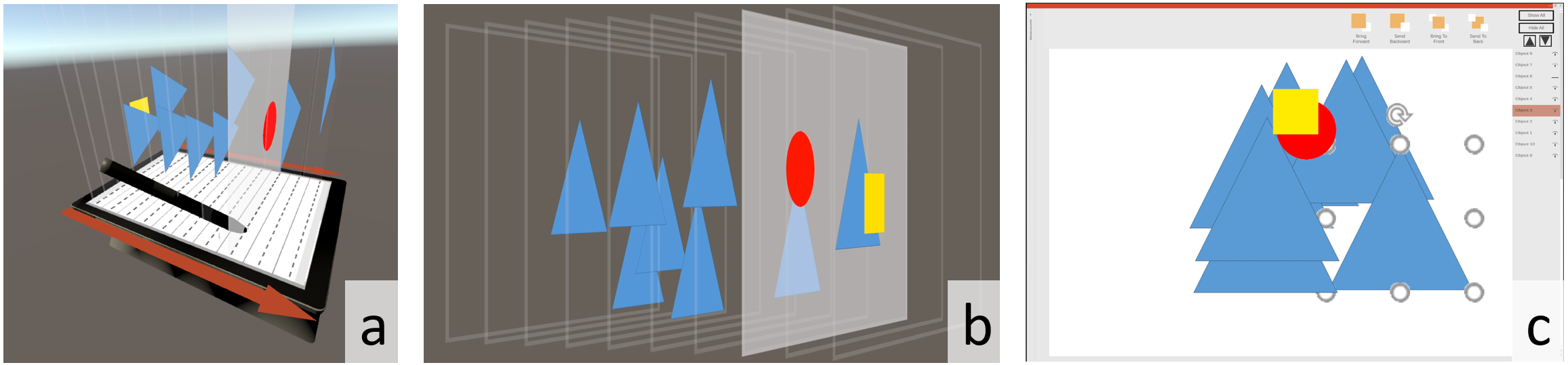}
	\caption{Conditions in the reordering task: a) our \textsc{VR} technique. b) a re-implementation of the \textsc{dynamic reordering} technique from PowerPoint for Mac c) a re-implementation of the standard \textsc{PowerPoint} technique. }
	\label{fig:ReorderingTaskConditions}
	\vspace{-0.3cm}
\end{figure}

This experiment had one independent variable, \textsc{interface}, with three levels.
First, the \textsc{VR} condition, a simplified version of the layer visualization and manipulation interface described in section \ref{sec:HandlingOcclusions} displays the layers in the air above the tablet (Figure \ref{fig:ReorderingTaskConditions}, a).
Second, the non-VR \textsc{dynamic reordering} technique available in PowerPoint for Mac was used (Figure \ref{fig:ReorderingTaskConditions}, b).
Third, the non-VR baseline technique available in \textsc{PowerPoint} was used. It presents the layers in a list next to the slide, where they can be dragged up and down, and buttons to bring a layer to the front or back (Figure \ref{fig:ReorderingTaskConditions}, c). The dependent variables for this experiment were task completion time, usability (System Usability Scale, SUS)~\cite{brooke1996sus}, workload by using NASA TLX (unweighted version)~\cite{hart1988development}, simulator sickness (SSQ)~\cite{kennedy1993simulator} and user preferences.

\subsubsection{Participants}
Fourteen volunteers (2 female, 12 male, mean age 27.1, $SD=4.2$) took part in the study. All but one had prior VR experience and normal or corrected to normal vision. 

\subsubsection{Apparatus}
We implemented all three conditions using Unity 2019.4, in order to facilitate the task design, to get consistent logging data for all three techniques and to avoid potential confounds due to different input and output devices. In the \textsc{VR} condition, the object layers were rotated 90 degrees and lifted up above the tablet, so that the edge of each layer just touched the display. Therefore, it is possible to move a layer by touching the intersection on the tablet and moving it around. For the \textsc{dynamic reordering} condition, the reordering feature from PowerPoint for Mac was re-implemented. The object layers were presented in 3D on the tablet and could be moved by touching a layer and dragging it to its new position. For the \textsc{PowerPoint} condition, we implemented an application resembling the PowerPoint interface for rearranging object layers. Re-implementing this interface allowed us also to exclude potential confounds, such as visual noise added by extraneous buttons and submenus not relevant for the experiment. The two interfaces showed the same content as the original would, just leaving blank functionalities that are not used in the study.

For the \textsc{VR} condition we used a HTC Vive Pro Eye in combination with a Microsoft Surface Pro 4 tablet and an AZLink stylus pen.
The pen, the tablet and the HMD were tracked via an OptiTrack motion-tracking system, using 4 Optitrack Prime 13 cameras placed above the user and two Prime 13W wide angle cameras placed closer to the user to support tracking for the stylus which was equiped with smaller markers.
The VIVE-based HMD tracking was overriden to prevent interference between VIVE lighthouses and OptiTrack cameras.
Optitrack was chosen for the study setup, because it provides very accurate tracking for the HMD, tablet and pen. Using Optitrack, we could utilize a normal lightweight pen in contrast to rather heavy trackable pens that are currently available. In future work this could also be implemented as a mobile prototype and be evaluated in-the-wild.
Both the pen and the tablet were visualized in the virtual environment using 3D models of the real objects.
The commodity PC running the VR application received touch inputs remotely from the Microsoft Surface via UDP.
The two non-VR conditions (\textsc{dynamic reordering} and \textsc{PowerPoint}) were run directly on the Microsoft Surface Pro 4 tablet.
The study setup can be seen in Figure \ref{fig:studySetup}, a and b.

\subsubsection{Procedure}
Upon arrival, the participants were first asked to sign a consent form and to fill out a demographic questionnaire. Then the participants started with one of the three conditions. The condition order was counterbalanced between the subjects to avoid effects due to fatigue or learning. With 14 participants, it was not possible to exactly counterbalance, but no significant order effects were detected. For each conditions the task was repeated 32 times. After each condition, participants completed the Simulator Sickness questionnaire~\cite{kennedy1993simulator}, the System Usability Scale~\cite{brooke1996sus} and the NASA TLX questionnaire~\cite{hart1988development}. After all conditions were successfully completed, participants filled out a questionnaire about their preferred technique. We then conducted a semi-structured interview to understand their choice and give them the opportunity to give comments. In addition, we recorded the task completion times for all conditions. On average, this study took 30 minutes. Volunteers did not receive a compensation for participating.

\subsubsection{Results}
Repeated measures analysis of variance (RM-ANOVA) was used to analyze task completion times, which were non-normal and therefore log transformed. For multiple comparisons Bonferroni adjustments were used at an initial significance level of $\alpha=0.05$. Aligned Rank Transform~\cite{wobbrock2011aligned} was used for subjective data and errors that are not normally distributed (or could not be normalized using log transform).

The main results of the reordering task are shown in Table \ref{tab:resultsTableReordering}. Due to logging errors, we lost 12 samples for the task completion time from the first participant in the reordering task, so the mean task completion time for this participant only consisted of 20 samples. Statistical significance tests showed that the task completion time was significantly influenced by \textsc{Interface} in such a way that the \textsc{VR} method ($M=4.51$s, $SD=2.46$) was significantly faster than both the \textsc{dynamic reordering} ($M=14.5$s, $SD=7.4$) and \textsc{PowerPoint} ($M=16.1$s, $SD=6.39$) methods. But no significant difference between \textsc{dynamic reordering} and \textsc{PowerPoint} was detected.

The NASA task load index also showed a significant influence of the \textsc{interface} on the overall task load.  Pairwise comparisons showed that the \textsc{VR} ($M=20.1$, $SD=11.46$) interface resulted in a significantly lower task load than the \textsc{dynamic reordering} ($M=49.29$, $SD=22.15$) interface. But no significant difference was detected between \textsc{VR} and \textsc{PowerPoint} ($M=34.23$, $SD=21.56$) or between \textsc{dynamic reordering} and \textsc{PowerPoint}. No significant differences between the \textsc{interface}s regarding simulator sickness were detected.

The usability was also significantly influenced by the \textsc{interface} in such a way that the usability of the \textsc{VR} method ($M=89.11$, $SD=7.76$) was significantly higher than for the \textsc{dynamic reordering} ($M=53.39$, $SD=21.63$) and \textsc{PowerPoint} ($M=69.64$, $SD=18.76$) methods. Again, no significant difference between \textsc{dynamic reordering} and \textsc{PowerPoint} could be detected.

All but one participant preferred the \textsc{VR} method. One preferred the \textsc{dynamic reordering} method. Six participants that preferred the \textsc{VR} techniques said that "it results in the fastest overview" (P1, P2, P3, P4, P7, P8) and "if something was occluded you could move your head" (P2, P7, P14). Five also mentioned that "it was easy to select the layers" (P1, P4, P6, P8, P12), three that "the interaction was more convenient" (P8, P9, P12,) and three that "it was easier and faster" (P1, P2, P11). Three participants also complained about the \textsc{dynamic reordering} condition. One said "I was confused which slide is selected" (P1), one that "it was hard to identify the layer" (P2) and one that "I often selected the wrong layer" (P6). Another one mentioned that "it is a problem that it is only displayed in 2D, so it is hard to see where to tap" (P12). One participant proposed to highlight the slides when touching them and selecting them by for example pressing the pens button.
This is in line with our observations that the two baseline techniques required a lot more trial and error to select the right objects, as they were partially occluded and it was hard to see which object belongs to which layer. In contrast, the three-dimensional view and head movement in VR helps with assigning objects to layers.

\begin{table}[t]
    \centering 
    \tiny
    \begingroup
    \setlength{\tabcolsep}{5pt}

    \begin{tabular}{|>{\centering}m{0.9cm}||>{\centering}m{0.3cm}|>{\centering}m{0.3cm}|c|c|c|}
        \multicolumn{6}{c}{\small\bfseries\textbf{Reordering Task}} \\

        \hline 
        & d$f_{1}$ & d$f_{2}$ & F & p &  $\eta^2_p$   \\ 
        \hline 

        TCT &  \cellcolor{lightgray}$2$ & \cellcolor{lightgray}$26$  & \cellcolor{lightgray}$71.8$  & \cellcolor{lightgray}$<0.001$  & \cellcolor{lightgray}$0.85$       \\
        \hline 

        TS-SS & $2$ & $26$ & $0.68$ & $0.52$ & $0.05$         \\
        \hline 

        SUS &  \cellcolor{lightgray}$2$ & \cellcolor{lightgray}$26$ & \cellcolor{lightgray}$22.8$ & \cellcolor{lightgray}$<0.001$ & \cellcolor{lightgray}$0.64$       \\
        \hline 

        Overall task load &  \cellcolor{lightgray}$2$  & \cellcolor{lightgray}$26$   & \cellcolor{lightgray}$18.98$   & \cellcolor{lightgray}$<0.001$   & \cellcolor{lightgray}$0.59$        \\
        \hline 
    \end{tabular}

    \endgroup

    \caption{RM-ANOVA results for the reordering task.  Gray rows show significant findings. TCT: Task Completion Time. TS-SS: Total Severity Dimension of the Simulator Sickness Questionnaire. SUS: System Usability Scale d$f_1$ = d$f_{effect}$ and d$f_2$ = d$f_{error}$.}
    \label{tab:resultsTableReordering}
    \vspace{-0.3cm}
\end{table}

\section{Usability Evaluation}
In addition to the performance evaluation that we presented in Section \ref{sec:PerformanceEvaluation}, we conducted a usability study on our prototype, which  consists of the techniques presented in Section \ref{sec:Concept}.
Our objective was to find out if they are easy to understand and to use for regular users.
Also, through participant's comments we gained valuable insights into how to improve our prototype. This study was divided into four parts representing the techniques from section \ref{sec:Concept}---object manipulation, handling occlusions, animations and working across slides. The three concepts (slide overview, multiple content sources, copying content) on working across slides, as presented in section \ref{sec:Concept}, were presented to the participants jointly as one coherent workflow. 

\subsection{Participants}
Eighteen participants (5 female, 13 male) took part in this study. Their mean age was $28.94$ years ($SD=5.3$). All had normal or corrected to normal vision and all but one had prior VR experience.

\subsection{Apparatus}
The VR applications described in this paper were implemented using the Unity 2019.4. We used a HTC Vive Pro Eye, which provides built-in eye tracking in combination with a Microsoft Surface Pro 4 tablet and an AZLink stylus pen. For the user study, the pen, tablet and HMD were tracked via an OptiTrack motion-tracking system with 6 Optitrack Prime 13 cameras (Figure \ref{fig:studySetup}, a and b), since pen tracking via VIVE-trackers or other VIVE-based tracking devices was unfeasible due to pen weight concerns. Therefore, the VIVE-based HMD tracking was overriden to prevent interference between VIVE lighthouses and OptiTrack cameras. Pen, tablet and the two fingers of the non-dominant hand, that were used for pinching, were visualized in the virtual environment, using 3D models of the real pen and tablet and spheres for the fingers. The tablet was connected to the PC running the VR application which received the touch inputs via UDP.

\subsection{Procedure}
First, participants were asked to sign a consent form and fill out a demographic questionnaire. Then the eye-tracking was calibrated using the built-in routine of the HTC Vive Pro Eye. 
All participants started with object manipulation, because it is a prerequisite for the animation techniques. 
The order of the remaining parts - handling occlusions, animations and working across slides - was counterbalanced. For each concept, the participants were walked through the possibilities and interaction techniques that are provided in the prototype. Then they had time to try out the technique as long as they liked.  On average participants spent about 5 minutes exploring each technique. Following each interaction concept, the participants orally graded three statements (while wearing the HMD) by giving a score on a seven-item Likert scale, regarding ease of use ("I would find the application easy to use"), utility ("I would find the application to be useful") and enjoyment ("I would have fun interacting with the application") (1: totally disagree, 7: totally agree). 
Also they were encouraged to think out loud about their experience and make suggestions. At the end, when all concepts were explored, the participants completed the Simulator Sickness questionnaire~\cite{kennedy1993simulator}, the System Usability Scale~\cite{brooke1996sus} and the NASA TLX questionnaire~\cite{hart1988development}.
Also, they were asked to rank the four techniques by popularity and we conducted a semi-structured interview to give the participants a chance to further express their thoughts.
The whole study took about 45 minutes on average.

\subsection{Results}

\begin{figure}[t]
	\centering 
	\includegraphics[width=0.7\columnwidth]{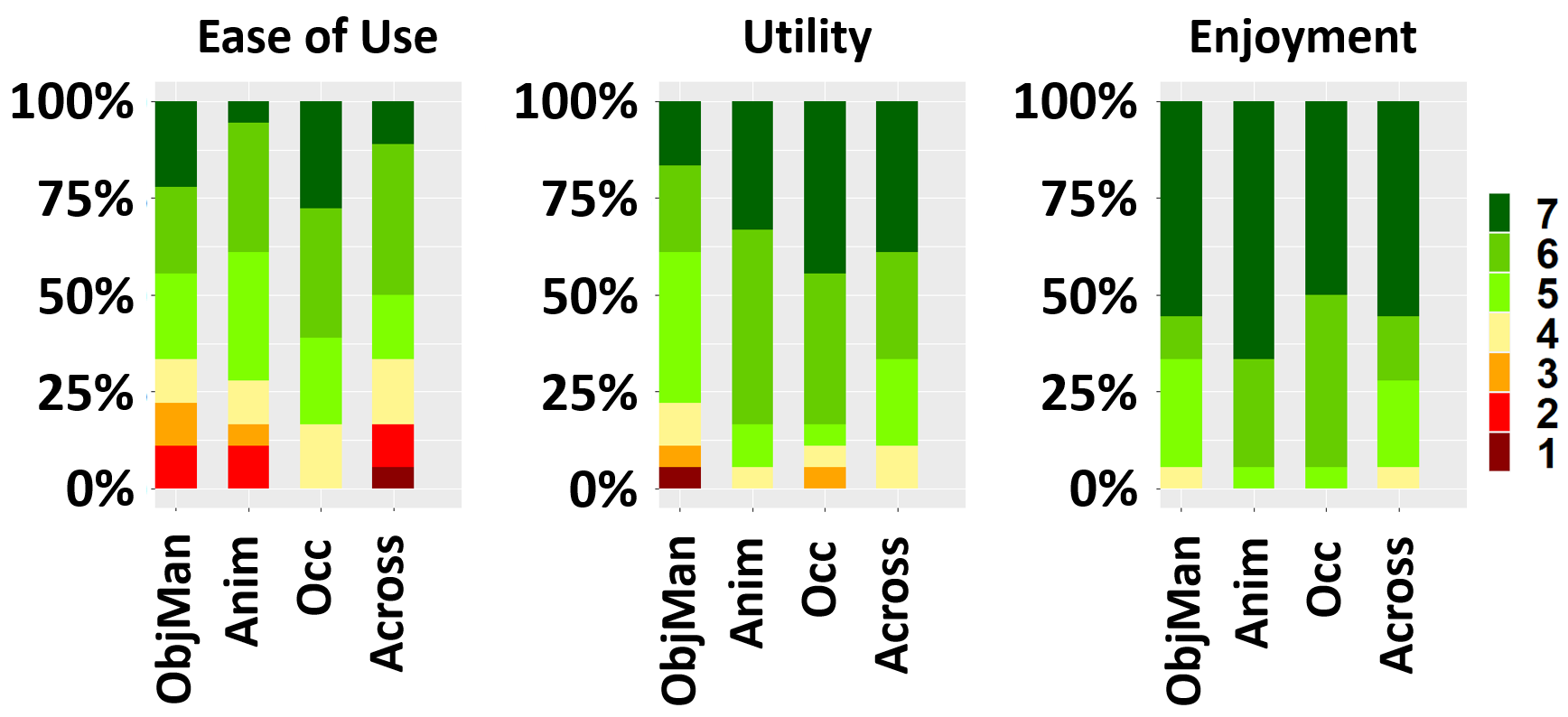}
	\caption{Answers from questionnaire about Ease of use, utility and enjoyment for the four concept with 7 being the highest and 1 the lowest possible score. ObjMan = Object Manipulation, Anim = Animation, Occ = Occlusion Handling, Across = Across Slides}
	\label{fig:ShortUsabilityQuestionnaire}
\end{figure}

The results from the three questions on ease of use, utility and enjoyment that were asked after each technique are presented in Figure \ref{fig:ShortUsabilityQuestionnaire}. It can be seen that the ratings for utility and enjoyment are high with more than 75\% of the answers being at least a five on the seven-item Likert scale. Even tough the participants had to learn a lot
in a short time, the ease of use rating was also high.

After each technique was presented, participants were asked to fill out three questionnaires (SUS, NASA TLX, SSQ).
The system usability questionnaire reported an average usability score of $71.53$ ($SD=20.13$) which indicates that our prototype has an average usability. This is an acceptable result for a prototype. 
The average total severity dimension of the Simulator Sickness Questionnaire was $12.26$ ($SD=17.34$) (Nausea: $M=5.83$, $SD=13.15$, Oculo-motor: $M=11.79$, $SD=19.49$, Disorientation: $M=15.47$, $SD=15.76$) , which means the participants were not suffering from severe simulator sickness. 

The average overall task load measured by the NASA task load index was $25.54$ ($SD=11.18$), indicating that they were not overwhelmed (due to technical problems, we lost the data for TLX of one participant and computed the average with one less value).

\begin{figure}[t]
	\centering 
	\includegraphics[width=0.6\columnwidth]{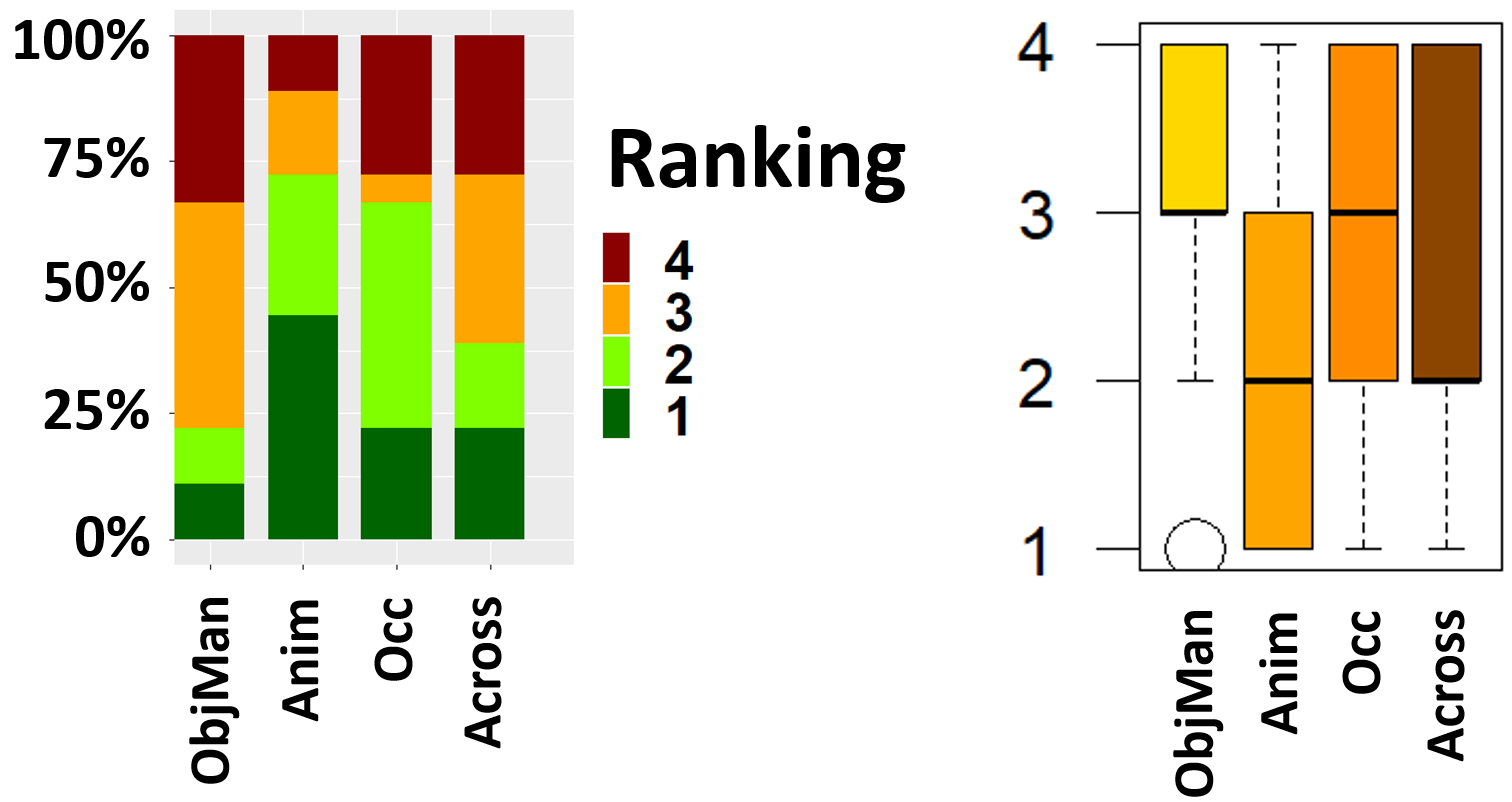}
	\caption{Left: percentage of people ranking the technique either 1st, 2nd, 3rd or 4th. Right: boxplot of the ranking. ObjMan = Object Manipulation, Anim = Animation, Occ = Occlusion handling, Across = Across Slides}
	\label{fig:rankingOfTechniques}
	\vspace{-0.3cm}
\end{figure}

Participants were also asked to rank the techniques based on preference, see Figure \ref{fig:rankingOfTechniques}. There is no clear trend visible, but animation and occlusion handling seem to be more popular than the object manipulation technique or working across slides.

Participants also had the chance to comment on the techniques. For the object manipulation technique, four participants mentioned that it requires some practice and P8 suggested an additional explicit mode switch.
P2 would rather do the scaling in 2D, as it only affects two dimensions. P7 thought the 3D rotation was intuitive, and P15 said it was very useful, because "you can better see what happens".

Regarding the animation, two participants mentioned that it is easier and more intuitive than in a standard slide authoring application. They also suggested additional features such as the snapping of two frames (P5), other types of animation (P8) and the possibility to exactly enter time (P16). However, P2 and P12 also mentioned that a certain amount of practice time is needed. 

P1 and P17 thought that the occlusion handling method makes it easier and more natural to find a specific layer. P8 suggested to add the possibility to see the final image not only on the right but also on the tablets surface or standing on the upper edge. To be more consistent with the animation mode, P4 suggested to move the layers in-air and P15 was concerned it could get confusing with a very large number of objects, but for standard slides this should not be an issue.

When working across slides, participants also mentioned that many new techniques need to be remembered (P1, P2, P12). Only one participant thought that too much content is displayed that confuses her. P2 liked the possibility to make the content in front of the user invisible and P16 would also like this feature for the slides presented to the sides of the tablet. P3 suggested to add the possibility to bring resources (like a PDF) closer, so that it is easier to read it. P15 found it challenging to hit the bezel while not looking at it. P10 mentioned to feel some motion-sickness when the content was moving around (when adding or closing content).

Participants were also asked to comment on the overall system and explained their choices for the rating. Four people said that they liked the system. P2, P10 and P14 said they had fun using the system. 
P6 and P17 could not imagine themselves using a VR HMD for work, but nevertheless, they liked the system.
Five participants mentioned that they would need to practice  to get used to all techniques.  However, P11 mentioned that "with a little practice it would be faster than in standard PowerPoint". P8 suggested to display a cheat sheet somewhere, as VR provides a lot of space.
Three participants mentioned that 3D view is useful and "makes things easier to see than in 2D" (P10). Four people mentioned the system seems viable and P14 said "I could imagine to use this system".
P15 and P17 liked the display of additional information and the increased workspace. P14 was interested in seeing how these interaction techniques could be integrated in a standard desktop setup.
P4, P17 and P18 especially liked the occlusion handling technique, because "it saves us a lot of work" (P4) and "it is what I always missed in PowerPoint" (P17).  
P14 and P15 especially liked the animation, because "it was fun" (P14) and "implemented well and easy to understand" (P15).

\section{Discussion}



As people work more remotely, from touchdown offices, on the go or from home, the importance of mobility and privacy increases. This paper, and other prior works \cite{Biener2020Breaking, Gesslein2020Pen} show that this mode of work can be more efficient than the current use of 2D displays, as HMDs can fill the user's full field of view with potentially very large virtual displays and show world grounded stereoscopic information, which the user can interact with directly.

To examine the effect of such a display on presentation authoring applications, we focused on four specific techniques in this paper: The use of the large field of view for selection and interactions with content catalogs, use of 3D visualization for animation, handling object ordering and occlusions, and  tracking the user's stylus in 6 DOF, enabling complex manipulations of objects.


We evaluated the performance of two techniques utilizing the larger display space and the 3D view provided by VR.
First, we examined the effect of the extended display space where we expected that a larger field of view would speed up a visual search task as indicated by prior work \cite{ragan2015effects}. The results of the search study showed that this was true for the \textsc{easy} conditions where the target was identifiable pre-attentively. Yet, contrary to our expectations, as matching required more mental effort per item, the field of view did not seem to significantly influence search time. Nevertheless, we argue that a wider field of view is desirable, as it performs at least as good or better when compared to a smaller field of view when identifying targets and has the potential to skip interactions for switching or toggling displayed information.

Our second performance study showed that VR-based 3D visualization for resolving occlusion outperformed two baseline techniques in terms of speed and usability. Follow-up interviews suggested that the technique was well received, as "easy", "fast" and "provided a good overview". Similar techniques for embedding 2D data in 3D could also be used in other applications from animation (representing time as the third dimension) or image editing (showing semantic layers and versioning in space) to displaying alignment constraints and relations between slide objects and more. We showed the benefits of VR as a work space, and hope to encourage more work in this direction.

Additionally, a usability study in a walk-up-and-use scenario was conducted to evaluate these techniques. Subjects were confronted with a large amount of new interaction techniques and input modalities that differ from traditional touch input techniques in several ways. In spite of this, participants gave positive ratings for ease of use, which is also reflected in the results of the system usability scale indicating an average usability of $71.53$. 
It showed the feasibility of our approach and the techniques were rated usable and enjoyable by participants. Many participants also expressed the feeling that their level of comfort with the techniques would improve with more training, indicating their unfamiliarity due to the walk-up-and-use scenario but general level of comfort with the techniques. Future work could look at ways to further improve the techniques to be even more intuitive, discoverable, and require less setup and explanation to use or to determine the actual learning curve of the techniques.
No severe levels of simulator sickness were measured among participants and their perceived task load was also not high, indicating that they had no major issues with the basic functionality of our prototype. We hope that these findings are precursors for supporting further tasks in presentation authoring beyond the tested ones. 
Our goal was to gain initial insights into the usability, yet it will be important for future work to evaluate them in a more extensive way.


In this paper we focus on the graphic organization of slides. Text entry is an important issue for presentation authoring, yet we did not address it in this paper, as prior works has already addressed typing in VR, e.g., \cite{grubert2018text, Dube2019}. These techniques can be used in conjunction with the presented techniques, for example, after entering the text, it could be manipulated like any other object (translated, rotated, scaled).


One of our main objectives in designing interactions was to use small hand movements, to allow for longer interaction times without fatigue. One option to further extend the input space but still keep small hand movements would be to remap physical pen and finger movements using a C-D ratio and visulizing copies of the pen and fingers (similar to our technique used when working across slides), for example to allow users to reach higher times in the 3D time view of the animation mode without scrolling through the timeline.

Another outcome from this work is the use of eye-gaze along with retargeting the input of a stylus on the tablet. The use of such techniques allows the user to interact with a very large display space while working in a limited cluttered physical environment, as their hands are located on a small tablet screen. This skips the need to physically reach displayed content sources.

Finally, this paper has only looked at the authoring side of presentation applications. There is another aspect of these applications which is the presentation process. While being out of the scope of this work, there are many similar advancements that can come to play while presenting, from the use of the large display space to presenting information useful for the presenter such as upcoming slides or notes. Also the presented techniques could be used to control the presentations such as quickly switching between slides that are not neighbours.

\section{Conclusions and Future Work}

In this work we prototyped an experience called PoVRPoint: a set of tools that couple pen-, touch- and gaze-based authoring of presentations on mobile devices with the interaction possibilities afforded by VR. We studied the utility of extended display space in VR for tasks such as visual search, spatial manipulation of shapes, animations and ordering overlapping shapes. The results showed that VR can improve usability and performance of common authoring tasks and are liked by participants.

We see multiple avenues of future work. First, we aim at investigating the knowledge worker experience within the office of the future in VR~\cite{grubert2018office}, with multiple applications in use at any given time. To achieve this, we plan to explore techniques for transferring content across applications. Second, we want to explore how to expand the knowledge worker experience in VR by opportunistically leveraging available physical objects \emph{in-situ}, such as a tray surface or an armrest in an airplane. 
It has been shown that VR can improve the usability and performance of editing presentations during the limited time period of the study. Still, future work should evaluate the effects of working in VR for prolonged time periods.
Finally, we would like to extend the work into a collaborative one, and see how we can further use the advantages of VR to create experiences with awareness of the context and the remote participants. For example, VR can enable remote collaboration that represents both the shared document (the task space), as well as a representation of collaborators with a reference space to show where they are pointing in relation to the shared document, and private spaces~\cite{tang2010ThreesCompany}. 



\bibliographystyle{abbrv-doi}

\bibliography{template}
\end{document}